\documentclass[]{aa}
 \usepackage{graphicx}

\begin{document}

   \title{Two-year monitoring of intra-day variability of quasar 1156+295 at 4.8~GHz}

   \author{B.-R. Liu\inst{1,3,4}
           \and
           X. Liu\inst{1,2}
           \and
           N. Marchili \inst{5,6}
            \and
           J. Liu\inst{1,2,3}
           \and
           L.-G. Mi\inst{1,3}
           \and
           T.P. Krichbaum \inst{6}
           \and
           L. Fuhrmann \inst{6}
           \and
           J.A. Zensus \inst{6}
           }
\offprints{X. Liu, liux@xao.ac.cn}

\institute{Xinjiang Astronomical Observatory, Chinese Academy of
Sciences, 150 Science 1-Street, Urumqi 830011, PR China \\
\email{liux@xao.ac.cn}\and Key Laboratory of Radio Astronomy, CAS,
Nanjing 210008, PR China \and University of Chinese Academy of
Sciences, Beijing 100049, PR China \and Department of Physics and
GXU-NAOC Center for Astrophysics and Space Sciences, Guangxi
University, Nanning 530004, PR China\and Dipartimento di Fisica e
Astronomia, Universit\`a di Padova, via Marzolo 8 I-35131 Padova,
Italy \and Max-Plank-Institut f\"ur Radioastronomie, Auf dem
H\"ugel 69, 53121 Bonn, Germany}

\date{Received / Accepted }

\abstract
  {}
  {The quasar 1156+295 (4C +29.45) is one of the targets in the Urumqi monitoring program which aimed to search for evidence of annual modulation in the
  timescales of intra-day variable (IDV) sources.}
   {The IDV observations of 1156+295 were carried out nearly monthly from October 2007 to October 2009, with the Urumqi 25m radio telescope at 4.8 GHz.}
  {The source has shown prominent IDV of total flux density in most observing sessions with variability timescales of $\leq$\,1 day
  at 4.8 GHz.
The estimated IDV timescales seem to follow an annual cycle that
can be fitted with an anisotropic interstellar scintillation (ISS)
model
  suggesting that a significant part of the flux density variations is due to ISS. The source underwent a dramatic flare in
  2008. We studied the possible consequences of the flare on the IDV of 1156+295 by comparing the changes in
  its IDV
  characteristics with the evolution of the 43 GHz Very Long Baseline Array (VLBA) core size of the source. The table 3 that contains
  all flux measurements is only available in electronic form
  at the CDS via anonymous ftp to cdsarc.u-strasbg.fr (130.79.128.5) or via
  http://cdsweb.u-strasbg.fr/cgi-bin/qcat?J/A+A/.}
 {The quasar 1156+295 shows evidence for an annual modulation of its IDV timescales at 4.8 GHz, the ISS-induced IDV timescales and
 variability strength might be affected
by the overall activity state of the source core. More frequent
IDV and VLBI measurements are required to confirm the relation
between the IDV appearance and the core-size evolution of the
source. }


   \maketitle

\section{Introduction}

\label{sec:intro}

Intra-day variability (IDV) has been observed in compact,
radio-loud active galactic nuclei (AGNs) over a wide range of the
electromagnetic spectrum. At centimeter wavelengths it was first
discovered in the mid-1980s (Witzel et al. 1986; Heeschen et al.
1987). Subsequent investigations showed that 25\% to 50\% of
flat-spectrum radio sources (Quirrenbach et al. 1992; Lovell et
al. 2008) and $\sim$60\% of bright {\it Fermi} blazars (Liu et al.
2011, 2012a) are radio intra-day variables. Both source-intrinsic
and source-extrinsic mechanisms have been proposed as possible
explanations of the variability (see Wagner \& Witzel 1995).
Source-intrinsic models attribute IDV to flux density variations
taking place directly in the AGNs. A simultaneous change in the
variability characteristics of radio and optical light curves
detected in the blazar S5~0716+714 during a four-week monitoring
campaign (see Quirrenbach et al. 1991) is the strongest argument
in favor of the possibility of source-intrinsic IDV at radio
bands. More recently, in a multifrequency campaign on the same
object, Fuhrmann et al. (2008) found stronger IDV at
short-centimeter and millimeter bands than at longer wavelengths,
confirming that the variability is probably intrinsic to the
source. The main issue with source-intrinsic explanations is that,
because of causality arguments, flux density variations on
timescales of less than one day imply very high brightness
temperatures for the emitting components (Qian et al. 1991),
violating the inverse-Compton limit $10^{12}$K (Kellermann \&
Pauliny-Toth 1969). Source-extrinsic models, which attribute IDV
to propagation effects, mainly due to interstellar scintillation
(ISS) in our Galaxy (e.g., Rickett et al. 1995; Dennett-Thorpe \&
de Bruyn 2000; Bignall et al. 2003; Rickett 2007), are less
affected by the brightness temperature problem.

In both source-intrinsic and -extrinsic models, the essential
characteristic of an IDV source is the very small size of the
emitting region. Since the light emitted by a very compact source
is expected to scintillate after crossing a plasma screen, the
idea that the turbulent interstellar medium may be the cause of
most of the detected IDV is not only reasonable, but almost
inevitable. An interesting implication of the standard ISS models
is that the timescale of the variability changes according to the
relative velocity between the Earth and the screen. Because of the
orbital motion of the Earth around the Sun, this velocity changes
across the year. If IDV is caused by ISS, the variability
timescales should undergo a one-year periodic cycle, commonly
referred to as the annual modulation of the IDV timescales.

The relationship which binds IDV to very small emitting regions
also explains how the IDV phenomenon reflects AGNs. All IDV
sources are flat-spectrum radio-loud AGNs (the most compact among
the AGNs), and almost all are blazars. Blazars are highly variable
through the whole electromagnetic spectrum, from the radio to
gamma-ray band; in the light of the AGN unified model (Urry \&
Padovani 1995), this is commonly interpreted in terms of the
beaming effect of a relativistic jet oriented very close to our
line of sight. The smaller the angle between the jet and the line
of sight, the smaller the apparent size of the emitting region.
Gamma-ray loud blazars are supposed to be more compact than
others, therefore a correlation between $\gamma$-ray emission and
IDV characteristics of blazars would naturally be expected.
Looking at the sources detected by the {\it Fermi} $\gamma$-ray
satellite, it appears that the IDV detection rate is higher in
{\it Fermi} blazars than in non-{\it Fermi} blazars (Liu et al.
2012a).

The arguments above suggest how strongly the IDV characteristics
of an AGN should relate to the structure of its emitting region.
Strong IDV was found in the visibilities of the VLBA data at 15
GHz of the quasar 1156+295 by Savolainen \& Kovalev (2008), and it
was explained with ISS. Furthermore, it is likely that the
emission of a new component in the jet of an IDV source leads to
an enlargement of the size of the scintillating region, inducing a
prolongation of the IDV timescale or even the quenching of the
IDV, as might have occurred in 0917+624 (Krichbaum et al. 2002;
Liu et al. 2013). Multiple scintillating regions related to the
appearance of new emitting components could be at work in some IDV
sources (e.g., in J1819+3845, Macquart \& de Bruyn 2007). These
variations in the structure of IDV sources may reflect on their
average flux density emission, as they are most likely associated
with strong outburst phases (Marchili et al. 2012).

Long-term monitoring of IDV sources is crucial for finding both
annual cycles in the IDV timescales, which could prove to be a
possible ISS origin of the IDV, and possible correlations between
the IDV characteristics and the flaring/quiescent state of a
source.

We have carried out a monitoring program for a sample of IDV
sources with the Urumqi 25m radio telescope at 4.8 GHz, aiming to
search for evidence of annual modulation in the timescales of
known IDV sources. At least three IDV sources (J1128+592, S4
0954+658, and S5 0716+714) in the monitoring program have shown
annual modulation patterns which favor an ISS origin of their
variability (Gab\'anyi et al. 2007; Marchili et al. 2012; Liu et
al. 2012b). The source 1156+295 (4C +29.45, with redshift of
0.724488, and Galactic coordinates l=199.41\degr, b=+78.37\degr)
is a flat-spectrum radio quasar, and also a {\it Fermi}
$\gamma$-ray source. It is one of the targets of our monitoring
campaign and was observed nearly on a monthly basis from October
2007 to October 2009. In this paper, we present the results of the
IDV observations of 1156+295, analyze the variability timescales
and discuss possible effects of the source flare that peaked
around March 2009 on the IDV.

\section{Observation and data reduction}

The Urumqi IDV observation campaign was carried out from August
2005 to January 2010 with the Urumqi 25m radio telescope, at a
frequency of 4.8 GHz. Observing sessions of three to five days
were typically scheduled once a month. The observing strategy and
the detailed calibration method were described in Marchili et al.
(2010); the accuracy of the calibrated data is of the order of
0.5\% in normal weather conditions. The quasar 1156+295 was
included in our monitoring program in late 2007, observed in 18
observing sessions from October 2007 to October 2009. One
observing session was too short ($<$\,1 day) to provide a
reasonable estimation of the variability characteristics,
therefore it was discarded. An example of a light curve for
1156+295 is shown in Fig. 1.

To quantify the variability of the source in each observing
session, we calculated the rms flux density, the rms flux density
over mean flux density (the so-called modulation index, $m$), and
the relative variability amplitude $Y$, defined as
$Y=3\sqrt{m^{2}-m_{0}^{2}}$, where $m_{0}$ is the mean modulation
index of all calibrators. This last parameter is meant to provide
an estimation of the true variability for the given observing
session, after removing the residual noise that still affects the
data. To evaluate the significance of the variability we used a
reduced chi-squared test. The whole procedure follows the strategy
proposed by Kraus et al. (2003).

The results of the statistical analysis for each observing session
are listed in Table~\ref{tab1}, along with basic information about
the observing sessions themselves. Table~\ref{tab1} also includes
the estimates of the variability timescales which are obtained
from the structure function (SF) introduced in the next section.
The columns give: (1) the epoch; (2) the day of year (DoY); (3)
and (4) the duration of observation and the effective number of
data points; (5) and (6) the IDV timescale and relative error from
the SF; (7), (8), and (9) the modulation index of calibrators, the
modulation index, and the relative variability amplitude $Y$ of
1156+295; (10) the source average flux density and the rms flux
density; (11) the reduced $\chi^2$ for the variability.

\begin{table*}
\centering
 \caption[]{The observational information and the results derived from the IDV observations at 4.8 GHz. }
\label{tab1}
         \begin{tabular}{ccccccccccc}
\hline
  \hline
    \noalign{\smallskip}
    1&2 &3 &4 & 5&6 &7 & 8&9 &10 &11  \\
 Start Day & DoY & dur& NP& $\tau_{sf}$ & error  & $m_{0}$ & $m$&  $Y$ & $\overline{S}_{4.8GHz}\pm rms$ & $\chi_r^2$ \\

    & &  (d)& &(d)&  & [\%] &
    [\%]& [\%] &  (Jy) & \\

\hline

  \noalign{\smallskip}

  13 Oct 2007  & 287    &  3.0    &  36      &  0.8   &  0.2   & 0.4& 2.4  & 7.1 &  1.022$\pm$0.024  & 17.12 \\

  21 Dec 2007  & 356     &  3.2    &  44      &  0.3   &  0.1    & 0.4&6.4 & 19.2 &  0.979$\pm$0.063  & 148.33 \\

  25 Feb 2008  & 57     &  2.9    &   28      &  0.3    & 0.1    &  0.6&3.5  & 10.3& 0.963$\pm$0.034  &  13.06 \\

 22 Mar 2008  & 82     &  3.0      &   39     &  0.4   &  0.1    & 0.4 &6.5  &  19.5 & 1.032$\pm$0.067  & 134.91 \\
 22 Apr 2008   & 113     &  3.1    &  36      &  0.4   &  0.1    & 0.5 &5.7  &17.0  &1.058$\pm$0.060  & 91.16 \\

 21 Jun 2008   & 173     & 3.5     & 63     &  0.4   &  0.1    & 0.5 &6.5  & 19.4 & 1.246$\pm$0.081    & 151.98 \\

  18 Jul 2008  & 200    & 4.8      &  44    &  0.3   &  0.1    &  0.6& 2.9  & 8.5 &1.417$\pm$0.041  & 16.85 \\

  20 Aug 2008  & 233     & 5.0      &  78    &  0.5   &  0.1    & 0.6 & 3.2   & 9.6 & 1.568$\pm$0.050   & 31.23 \\

 13 Sep 2008  & 257     & 3.0     & 45    &  0.8   &  0.2    & 0.4 & 3.9  &  11.6 &1.685$\pm$0.066  & 51.91\\

 6 Nov 2008  & 311    &  3.6      &   26   & 0.3   &  0.1    &  0.6 & 1.7 &  4.8 &2.074$\pm$0.035  & 5.06 \\


22 Dec 2008  & 357     & 2.3      &  26     &  0.4   &  0.1   & 0.4 &1.2  & 3.4& 2.428$\pm$0.028  & 3.67 \\


  11 Jan 2009  & 12    & 2.6      &  38    &  0.5   & 0.1    &  0.4 &3.4 &  10.1 &  2.628$\pm$0.090   & 33.67 \\

  21 Mar 2009  & 81    & 4.9     & 63      &  0.4  &  0.1  & 0.4 &1.9  & 5.6 & 2.740$\pm$0.051  & 10.90 \\

 19 Apr 2009  & 110   & 5.4    & 54       & 0.3   &  0.1    &  0.5 &1.3 &3.6  &2.629$\pm$0.033  & 4.43\\

  25 Jun 2009 & 177   & 2.6    & 26       &0.5   &  0.1   & 0.6 &1.4 & 3.8 & 2.382$\pm$0.034  & 5.93 \\

  22 Sep 2009  & 266 & 5.5    & 54      & -    &  -     & 0.6  & 0.8 & 1.6 &  1.817$\pm$0.014  & 1.60  \\

 9 Oct 2009  & 283   & 2.3   & 30       & 0.5    &  0.1   &  0.4 & 1.1 & 3.1  & 1.745$\pm$0.019 & 2.85 \\

           \noalign{\smallskip}
            \hline
           \end{tabular}{}
   \end{table*}

\section{IDV analysis and ISS model-fit}

With the chi-squared test of the $\chi_{r}^{2}$ reported in
Table~\ref{tab1}, the quasar 1156+295 shows IDV in most of the
observing sessions at a confidence level of $\geq\,99.9$\,\%,
except for the session of September 2009. The variability
amplitude $Y$ varies from about 3\% to 20\%, while the
peak-to-trough variations in the light curves, in some cases, are
as high as 25\% (see Fig.~\ref{fig1}).

\subsection{IDV timescale estimate}

The characteristic timescale of the IDV in 1156+295 was estimated
using the first-order SF analysis (Simonetti et al. 1985). The
value returned by the SF for a given time-lag $\tau$ is
proportional to the variance of the signal calculated using all
the pairs of data-points with time-separation $\sim \tau$. If the
variability of a signal has a characteristic timescale
$\tau_{sf}$, its variance should fluctuate around a constant value
for any $\tau > \tau_{sf}$. Above the noise level, the SF rises
monotonically with a power-law shape, reaching a plateau at the
characteristic timescale $\tau_{sf}$. Sometimes the structure
function may show more than one plateau, indicating the existence
of multiple variability timescales. In these cases, we identified
the characteristic timescale with the shortest plateau (Marchili
et al. 2012). The error of the IDV timescale is estimated by
considering the uncertainties in the evaluation of both the SF
saturation level and the power-law fit; generally, longer
timescales are affected by relatively larger errors.

\begin{figure}
     \includegraphics[height=6cm,width=8cm]{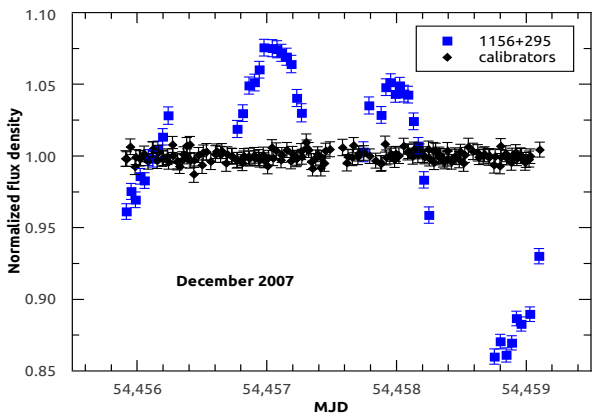}
     \caption{Normalized light curves of quasar 1156+295 (square) and calibrators (diamond) in December 2007 at 4.8 GHz. }
      \label{fig1}
\end{figure}

\begin{figure}
     \includegraphics[height=6cm,width=8cm]{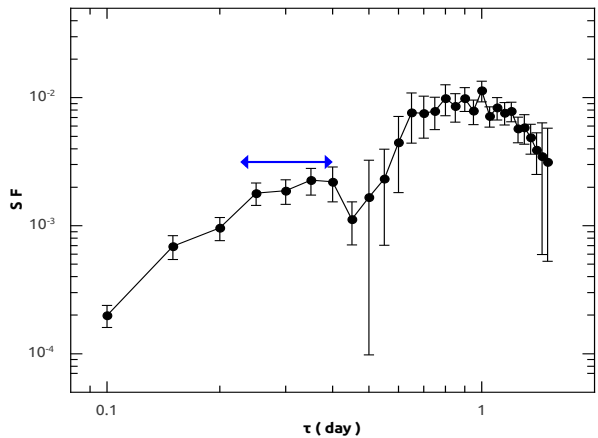}
     \caption{Structure function plot for the 2007 December data of 1156+295 at 4.8 GHz.}
      \label{fig2}
\end{figure}

In Fig.~\ref{fig2}, we provide an example of a structure function
for the light curve in December 2007 (Fig.~\ref{fig1}), with the
arrow marking the timescale area. The resulting IDV timescales
range from $\sim$\,0.3 to 1.0 day, typically around 0.4$\pm$0.1
day as listed in Table~\ref{tab1}.

\begin{figure}
     \includegraphics[height=6cm,width=8cm]{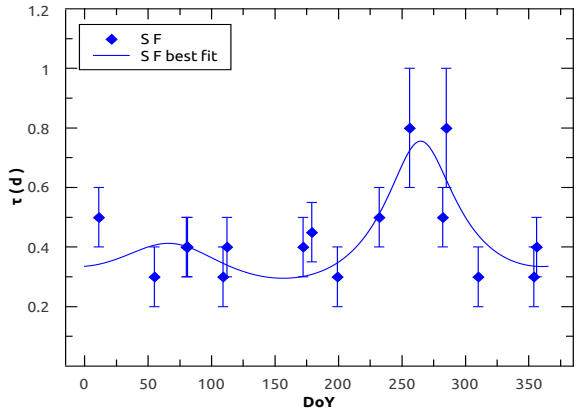}
     \caption{Annual modulation plot of the IDV of 1156+295 at 4.8 GHz, with the best-fit anisotropic ISS model
     for the IDV timescale.
     }
     \label{fig3}
   \end{figure}

    \begin{figure}
     \includegraphics[width=8cm]{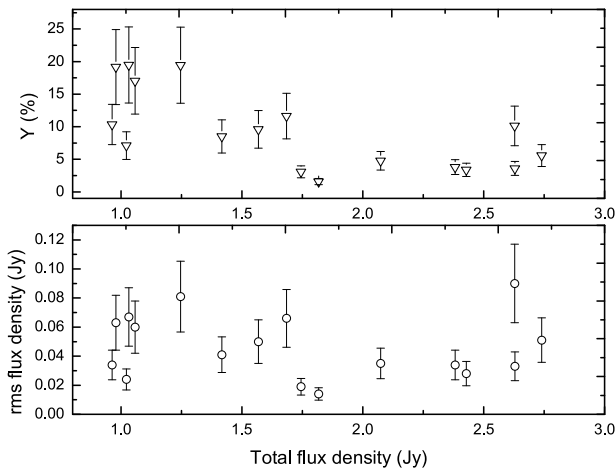}
     \caption{Relative variability amplitude $Y$ (upper panel) and rms flux density (lower panel) versus total flux density at 4.8 GHz.}
      \label{fig4}
   \end{figure}

\begin{figure}
     \includegraphics[width=8cm]{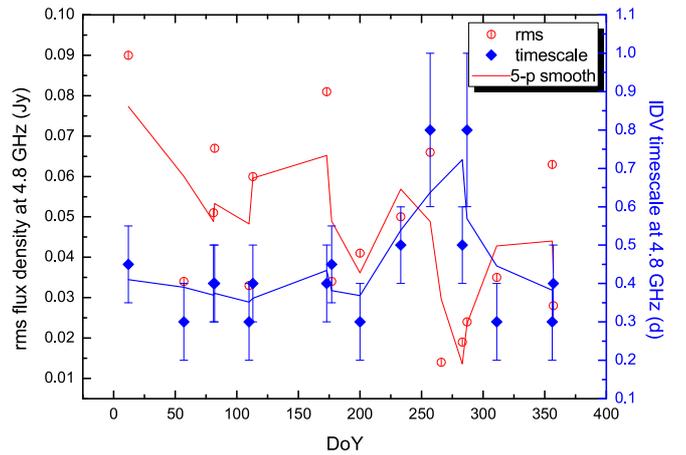}
     \caption{The rms flux density and the IDV timescale at 4.8 GHz versus day of year.}
      \label{fig5}
   \end{figure}

\begin{figure}
     \includegraphics[width=8cm]{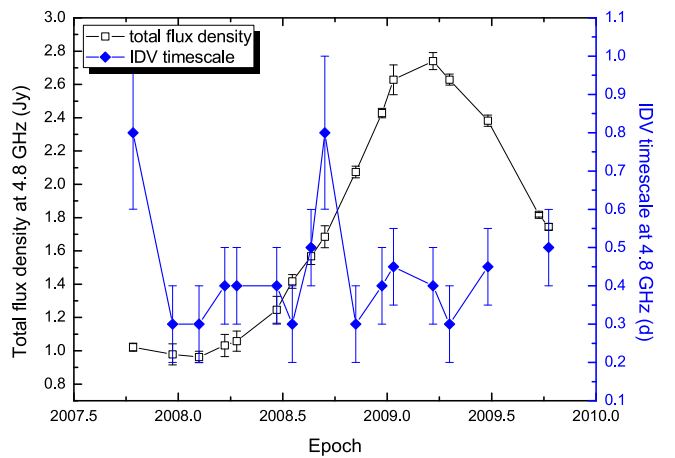}
     \caption{Total flux density and the IDV timescale at 4.8 GHz versus observing epoch.}
      \label{fig6}
   \end{figure}

\subsection{Annual modulation model fit}

To investigate the possible existence of an annual modulation in
the timescales, we fitted a model that relies on the formalism
introduced by Coles and Kaufman (1978), developed in Qian \& Zhang
(2001), and updated to take into account the case of anisotropic
scattering (Bignall et al. 2006; Gab\'anyi et al. 2007; see
Marchili et al. 2012 for more details). Figure~\ref{fig3} displays
the IDV timescales from the SF method versus day of year.
According to the anisotropic ISS model, the variation of the IDV
as a function of DoY depends on the orientation of the elliptical
scintillation pattern (described by the vector ${\bf
\hat{s}}=(\mathrm{cos\theta, sin\theta})$), the relative velocity
between the scattering screen and the Earth $\bf
v\mathrm{(DoY)}=\bf v_{\mathrm{ISS}}-\bf
v_{\oplus}\mathrm{(DoY)}$, the distance to the screen D, the
source's angular size $\theta_s$, and the anisotropy factor r:

\begin{equation}
\tau(\mathrm{DoY}) = \frac{\theta_s \mathrm{D}\cdot
\sqrt{\mathrm{r}}}{\sqrt{{\mathrm{v}^2(\mathrm{DoY})+(\mathrm{r}^2-1)\,({\bf
v(\mathrm{DoY}) \times \hat{s}})^2}}}.
\end{equation}

The algorithm for the least-squares fitting of the time scales
uses five free parameters: the relative velocity between the
scattering screen and the Earth ${\bf v}$ projected onto the right
ascension and the declination coordinates which allows one to fit
the screen velocity ($\bf v_{\mathrm{ISS},\alpha}$ and $\bf
v_{\mathrm{ISS},\delta}$) relative to the local standard of rest
(LSR), since the Earth's velocity is known with respect to the
LSR; the distance to the screen; the anisotropy degree; and the
anisotropy angle $\theta$ (measured from east through north),
which is derived from the vectorial product ${\bf v(\mathrm{DoY})
\times \hat{s}}$. We note that the distance cannot be
unambiguously calculated, unless the source angular size is known.
For a rough estimation of D, we assumed as an educated guess
$\theta_s=70 \mu as$.

The parameters that best fit the timescales of 1156+295, along
with the respective reduced chi-squared values, are reported in
Table~\ref{tab2}, and the anisotropic ISS model fitted curve is
shown in Fig.~\ref{fig3} for the IDV timescales. The IDV
timescales show evidence in favor of a seasonal cycle, with a
remarkable slow-down between DoY 250 and 290. The screen has a
relatively high velocity (with respect to LSR), at a distance of
0.17$\pm$0.04 kpc. The screen appears strongly anisotropic, with
an anisotropy ratio of 10.0$\pm1.0$ and an anisotropy angle of
170\degr$\pm$10\degr (measured from east through north). Such
highly anisotropic scattering is not rare; the three known fast
(intra-hour) scintillating sources all show evidence of highly
anisotropic scintillation patterns (i.e., quasar PKS 0405$-$385,
Rickett et al. 2002; quasar J1819+3845, Dennett-Thorpe \& de Bruyn
2003; quasar PKS 1257$-$326, Bignall et al. 2006). In our case,
quasar 1156+295 has a typical timescale of $\sim$\,0.4 day,
placing it between the intra-hour variables and the classical IDV
sources. It is claimed that in the one-dimensional scintillation
model proposed by Walker et al. (2009) an anisotropy degree can be
as high as $10^{5}$. For high Galactic-latitude scintillating
sources like 1156+295, the highly anisotropic plasma turbulence
may be concentrated in relatively thin layers lying along the
boundary of clouds (Linsky et al. 2008), for example.

We also tried to fit the IDV timescales with an isotropic ISS
model; it returns a lower velocity of the scattering screen
$v_{\alpha}=-$70$\pm30$ km/s and $v_{\delta}=$25$\pm15$ km/s, and
a larger distance of 0.29$\pm0.12$ kpc to the screen, as listed in
Table~\ref{tab2}. Given the lower number of free parameters, the
isotropic fit to the timescale provides a larger ${\chi_r}^{2}$
than the anisotropic model. In the following, we will focus on the
anisotropic ISS model result.

\begin{table}
         \caption[]{Screen parameters estimated from anisotropic and isotropic ISS models for the IDV timescales of 1156+295 at 4.8 GHz. }
         \label{tab2}
\tiny
\renewcommand\baselinestretch{1.18}
      \tabcolsep 0.7mm
         $$
         \begin{tabular}{c|c|c|c|c|c|c }

            \hline
            \noalign{\smallskip}

Method &$v_{\mathrm{ISS},\alpha}$ &$v_{\mathrm{ISS},\delta}$&Screen&Anisotropy& Anisotropy  & ${\chi_r}^2$\\
 &to LSR &to LSR & distance & degree & angle & \\
  & (km/s)& (km/s) & (kpc) & (ratio) & $\theta(E \rightarrow N)$  & \\
            \noalign{\smallskip}
            \hline
            \noalign{\smallskip}

Anisotropic& $-100\pm20$  &$30\pm15$ & $0.17\pm0.04$ & $10.0\pm1.0$ & $170\degr\pm10\degr$  & 1.1  \\

Isotropic& $-70\pm30$  &$25\pm15$ & $0.29\pm0.12$ & &    &2.5  \\
           \noalign{\smallskip}
            \hline
           \end{tabular}{}
         $$
   \end{table}

Both the modulation index $m$ and the relative amplitude $Y$  (see
Table~\ref{tab1}) follow a decreasing trend in the two years of
our monitoring campaign; by definition, the two quantities are
inversely proportional to total flux density. Their variations may
be caused by significant changes either in the rms or in the
average flux density which varies considerably because of a strong
flare peaking around March 2009 (see, e.g., the relative amplitude
$Y$ versus total flux density in Fig.~\ref{fig4}, upper panel).
Plotting the rms versus average flux density (Fig.~\ref{fig4},
lower panel), we do not find any significant correlation between
the rms and total flux density. It should be noted that for a
given light curve intended as a series of measurements of a given
signal, the rms value is influenced both by the characteristics of
the signal and by the sampling. The estimated rms of the same
signal may vary considerably depending on how densely and
uniformly the signal is sampled. We tried to estimate the
uncertainty in the reported rms values of the 1156+295 light
curves using two independent approaches: 1. In June 2008, the
Urumqi observations were performed quasi-simultaneously with the
Effelsberg observations. The length and the sampling of the
resulting light-curves are significantly different; they can be
considered as independent realizations of the same process. Their
rms values differ by $\sim$30\%. 2. We calculated the running
standard deviations over the longest 1156+295 light curves; the
time windows were chosen to be comparable to or longer than the
variability timescales, to ensure that a significant amount of
variability would not be lost. We compared the results obtained
for windows with different sizes and samplings, finding a
deviation from the average by about 25\%. These tests suggest that
it is reasonable to consider an uncertainty in the estimated rms
values around 30\%.

The rms and the IDV timescale are plotted versus DoY in
Fig.~\ref{fig5}. It appears that the longest timescale lies around
the lowest rms flux density, similar to the case of S5 0716+714
(Liu et al. 2012b), implying that the IDV amplitude may be lowest
around the timescale that is the slowest in the annual modulation.

 \begin{figure}
    \includegraphics[width=8.5cm]{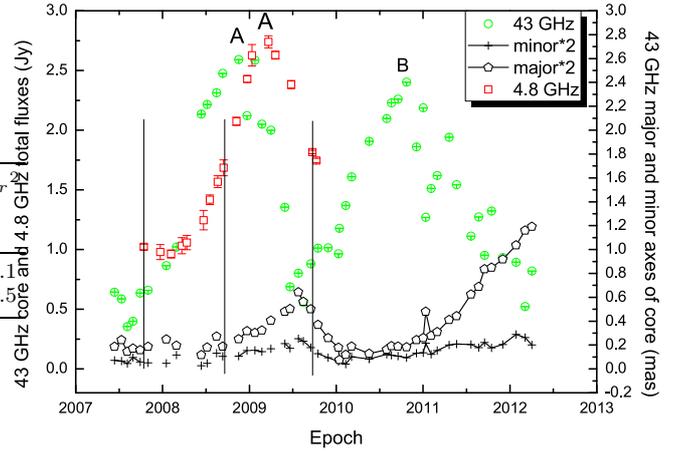}
    \caption{Integrated flux density values (green), the size of the major (times 2 for better display) and minor (times 2) axes of the 43 GHz VLBA core,
    and the 4.8 GHz total flux density (red) versus observing epoch. The vertical lines indicate the longer IDV timescales and the possible quenching of the variability}
    (the last vertical line); the letters A and B mark the flares of 1156+295.
     \label{fig7}
  \end{figure}

\section{Discussion of the source flare}

As can be seen in Fig.~\ref{fig6}, the flux density of 1156+295
underwent a significant flare. The flare appears to start in early
2008 at 4.8 GHz and the flux density monotonically increases to a
peak at early 2009. The peak-to-trough variation of flux density
is $\sim$\,170\%; the long-term variations should be due to a
source intrinsic property.

Lovell et al. (2003) first detected the IDV of 1156+295 in three
nights with the VLA at 5 GHz, and Lovell et al. (2008) reported
the modulation index $m$ of the IDV in this source of 5.8\%,
3.4\%, 4.6\%, and 4.5\% (with total flux density of 2.89 Jy, 2.89
Jy, 2.28 Jy, and 1.90 Jy, respectively) for four epochs from 2002
to early 2003 at 5 GHz in the MASIV project. Savolainen \& Kovalev
(2008) found a significant IDV as high as 40\% from peak to trough
with a timescale of 2.7 hours in the correlated flux density of
1156+295 in a VLBA observation at 15 GHz on February 5, 2007, and
the relative amplitude of the variations is larger than that
observed by Lovell et al. (2003) at 5 GHz. Savolainen \& Kovalev
(2008) argued that this fast IDV at 15 GHz must be due to ISS. In
a multi-wavelength campaign, a slower IDV in S5 0716+714 was
observed at 10.45 GHz, it has been mainly attributed to a source
intrinsic property (Furhmann et al. 2008). At 4.8 GHz in our case,
an annual modulation of IDV timescales is obtained for the first
time, suggesting that the IDV of 1156+295 at 4.8 GHz is dominated
by ISS.

High resolution VLBI images of 1156+295 at 15 GHz have revealed an
extremely core dominated jet (with a core fraction of $>$\,90\%),
with the largest apparent jet speed of 24.74c corresponding to
607$\pm$45\,$\mu$as/year (Lister et al. 2009). The inner-jet
structures of 1156+295 at 15, 43, and 86 GHz have also been
studied with the VLBA observations by Zhao et al. (2011). It is
possible that the source structure changes influence the
ISS-induced IDV timescale, e.g., the inner-jet position angle
changes lead to changes of the projected size of the scintillating
component as discussed in Liu et al. (2012b). It is also possible
that a new jet leads to a size of the core region (a blend of core
and the new jet) close to or larger than the scattering size
(i.e., the Fresnel angular scale, see Narayan 1992; Walker 1998),
so the IDV timescale could be prolonged, and the scintillation
could be quenched, leading to a decrease in the amplitude of the
variations, as might have occurred in 0917+624 (Krichbaum et al
2002), for example. After a few months or more, the new jet
component moves farther down the jet, expands, and fades, the
compact VLBI core becomes dominant again, and the normal
scintillation timescale resumes.

We studied the 2008 flare of 1156+295 in some detail, trying to
verify its possible effect on the ISS-induced IDV. To investigate
the evolution of the source structure, we have tried to model-fit
the 43 GHz VLBA core of 1156+295 with two-dimensional elliptical
Gaussian model from the Boston University 43 GHz VLBA monitoring
data\footnote{http://www.bu.edu/blazars/VLBAproject.html}(see
Marscher et al. 2012). The result is presented in Fig.~\ref{fig7}
(the relative errors of the model-fitted parameters are
$\sim$10\%), the model fitted core flux density shows that the
2008 flare of 43~GHz core in 1156+295 starts earlier at 43 GHz
than the total flux density flare at 4.8 GHz, and a time delay of
about five months is evident between the first peak A at 43 GHz
and the peak A at 4.8 GHz. The position angle of the model-fitted
major axis of the 43 GHz VLBA core ranges from $-10^{\circ}$ to
$+20^{\circ}$ with respect to the north; it is not aligned with
the anisotropic angle ($\sim$\,$-80^{\circ}$ from north) in
Table~\ref{tab2}. The major and minor axes of the core resulted
from the elliptical Gaussian modelfit are in Fig.~\ref{fig7}
(multiplied by a factor of 2 for a better display). The major axis
is quite stable in the rising phase of the 2008 flare A at 43 GHz,
but increases gradually from the down-phase of the flare at 43
GHz. There is an anti-correlation between the major axis and the
integrated flux density (coefficient -0.38, with the null
hypothesis probability $p$= 0.01) of the core component at 43 GHz.
The median value of the major/minor axis ratio is $\sim$2,
suggesting the model-fitted inner-jet could not fully account for
the high anisotropy degree of the scintillation in Table 2. The
minor axis appears to follow a similar trend of the variability of
the major axis, with a correlation between the two (coefficient
0.82, with the null hypothesis probability $p$= 8.0E-13). There is
no apparent correlation either between the integrated flux density
and position angle of the major axis or between the major axis and
position angle of the major axis. Statistically, the mean,
standard deviation, minimum, median, and maximum value of the
minor axes is 68, 33, 13, 65, and 144 in $\mu$as respectively. The
mean and median sizes of the minor axes at 43 GHz are comparable
to the assumed scintillating component size (70\,$\mu$as as
mentioned above) in the ISS modelfit at 4.8 GHz. We see that the
longer IDV timescales in October 2007 and September 2008 (first
two vertical lines in Fig.~\ref{fig7}) are in the area of
relatively small core size, whereas September 2009 (the last
vertical line) is in the area of large core size.

\begin{figure}
    \includegraphics[width=8cm]{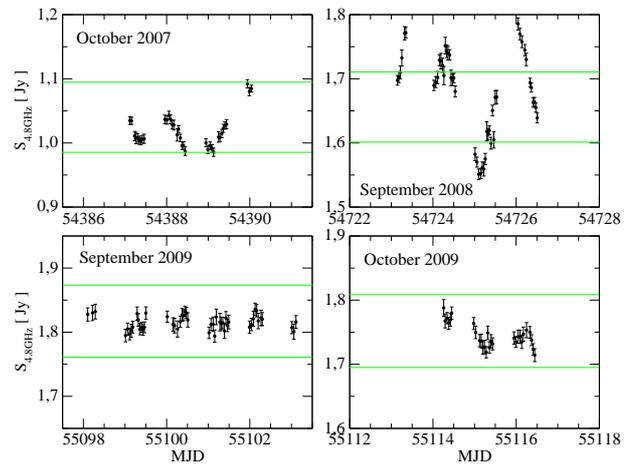}
    \caption{The light curves of 1156+295 during the slowdown phase of the annual modulation
    cycle. The green lines show the amplitude of the peak-to-peak variation in October 2007.}
     \label{fig8}
  \end{figure}

In Fig.~\ref{fig6}, no linear correlation (coefficient -0.11, with
the null hypothesis, i.e., no correlation probability $p$= 0.68)
is found between the IDV timescale and the total flux density,
implying that the source flare may have not generally influenced
the IDV timescale. The September and October 2009 observations,
however, show some characteristics that strongly differentiate
them from any previous one. In Sect 3.2, we mentioned that the
slowdown phase (between September and October) in the annual
modulation of the 1156+295 timescales coincides with a decrease of
the rms of the measured flux density. As shown in Fig.~\ref{fig8},
the amplitude of the variations observed in September and October
2009 (lower panels) is remarkably lower than in October 2007 and
September 2008 (upper panels). In September 2009 the measured
variability is comparable with the residual noise in the
calibrators, so that it is impossible to estimate a variability
timescale. This could be the consequence of a characteristic
timescale that is considerably longer than the ones estimated in
the same period of previous years, and/or the effect of a
quenching of the scattering. In October 2009 the variations in the
flux density are visible again, indicating that the variability
timescale is comparable to the observation duration, but the
amplitude of the variations is still lower than in previous years.

Considering the time delay between the peaks of the flare at 43
and 4.8 GHz, we may hypothesize that the source's angular size at
4.8 GHz could reach its peak a few months later than at 43 GHz,
probably around the end of 2009. It is therefore possible that the
peculiarities of the variability characteristics of 1156+295 in
September and October 2009 have to do with structural changes in
the source's emitting region.

\section{Summary}

We have carried out nearly monthly IDV observations of quasar
1156+295 in two years, with the Urumqi 25m radio telescope at 4.8
GHz, and the source has shown prominent IDV in most observing
sessions. We estimate the IDV timescale with the method of
structure function analysis. The IDV timescales stacking through
the day of year can be best-fitted with an anisotropic ISS model,
and although the data are relatively sparse, the result is
positive in favor of an annual modulation of the IDV timescales,
suggesting at least a significant part of the flux density
variations are due to ISS. The source underwent a dramatic flare
peaking around March 2009 at 4.8 GHz. We analyzed the flux density
and size evolution of the model-fitted VLBA core component at 43
GHz to find possible influences on the IDV of the source. We
hypothesize that the very low variability amplitudes measured in
September and October 2009 may be the consequence of the increased
source size related to the flare.

\begin{acknowledgements}

We thank the anonymous referee for insightful comments that have
improved the paper. This work is supported by the 973 Program of
China (2009CB824800) and the National Natural Science Foundation
of China (Grant No.11073036, No.11273050, No.11203008). N.M. is
funded by an ASI fellowship under contract number I/005/11/0. This
study makes use of 43 GHz VLBA data from the Boston University
gamma-ray blazar monitoring program
(http://www.bu.edu/blazars/VLBAproject.html), funded by NASA
through the Fermi Guest Investigator Program.

\end{acknowledgements}

\end{document}